\documentclass[11pt]{article}
\usepackage{moriond,epsfig}

\def\Journal#1#2#3#4{{#1} {\bf #2}, #3 (#4)}

\def\NPB{{\em Nucl. Phys.} B}

\def\PRD{{\em Phys. Rev.} D}
\def\ZPC{{\em Z. Phys.} C}

\def\st{\scriptstyle}

\def\ra{\rightarrow}

\def\be{\begin{equation}}
\def\ee{\end{equation}}
\def\bea{\begin{eqnarray}}
\def\eea{\end{eqnarray}}

\def\greaterthansquiggle{\raise.3ex\hbox{$>$\kern-.75em\lower1ex\hbox{$\sim$}}}
\def\lessthansquiggle{\raise.3ex\hbox{$<$\kern-.75em\lower1ex\hbox{$\sim$}}}
\newcommand{\beq}{\begin{equation}}
\newcommand{\eeq}{\end{equation}}
\newcommand{\beqa}{\begin{eqnarray}}
\newcommand{\eeqa}{\end{eqnarray}}
\newcommand{\ba}{\begin{array}}
\newcommand{\ea}{\end{array}}

\newcommand{\cP}{{\cal P}}

\def\s               {\sigma}

\def\D               {\Delta}

\def\ti    {\tilde}

\def\st    {{\ti t}}

\def\cth   {\cos\theta}

\def\cst   {\cos\theta_{\ti t}}

\newcommand{\mst}[1]   {m_{\ti t_{#1} }}

\def\Pm  {{\cal P}_-^{}}
\def\Pp  {{\cal P}_+^{}}
\def\fbi {{\rm fb}^{-1}}

\begin{document}  
\vspace*{4cm}
\title{Production of Charginos, Neutralinos, and Third Generation 
Sfermions at an {\boldmath $e^+e^-$}~Linear Collider}

\author{A.~Bartl,$^1$~ 
H.~Eberl,$^2$~ 
H.~Fraas,$^3$~
S.~Kraml,$^2$~ 
W.~Majerotto,$^2$~
G.~Moortgat--Pick,$^4$~
W.~Porod\,$^5$}

\address{$^1$~Institut f\"ur Theoretische Physik, Universit\"at Wien, 
     A--1090 Vienna, Austria \\
$^2$~Inst. f. Hochenergiephysik,  
     \"Osterr. Akademie d. Wissenschaften, 
     A--1050 Vienna, Austria \\
$^3$~Institut f\"ur Theoretische Physik, Universit\"at
     W\"urzburg, D--97074 W\"urzburg, Germany \\
$^4$~DESY, Deutsches Elektronen--Synchrotron, D--22603 Hamburg,
     Germany \\
$^5$~Inst.~de F\'\i sica Corpuscular (IFIC), CSIC, 
     E--46071 Val\`encia, Spain}

\maketitle\abstracts{
We discuss the production of neutralinos, charginos, and third
generation sfermions in $e^+e^-$ annihilation 
in the energy range $\sqrt{s} = 0.2-1$~TeV. We present numerical
predictions within the Minimal Supersymmetric Standard
Model for the cross sections and
study the importance of beam polarization for the
determination of the underlying SUSY parameters.}

\section{Introduction}
The search for supersymmetric (SUSY) particles will be one of
the main goals of a future $e^+e^-$ linear collider 
with an energy range $\sqrt{s} = 0.5 - 1$~TeV. Such an $e^+e^-$
linear collider will also be very well suited for the
precision determination of the parameters of the underlying SUSY
model. This will be necessary to find out the 
mechanisms of SUSY breaking and electroweak symmetry breaking.

In this contribution we summarize the results of our recent
phenomenological studies on the production of charginos,
neutralinos and third generation sfermions in 
$e^+e^-$ annihilation at energies between $\sqrt{s} = 200$~GeV
and $1$~TeV \cite{chne,sf}. We consider the effects
of both $e^-$ and $e^+$ beam polarizations. 
Polarizing both beams can be advantageous for 
three reasons: (i) One can obtain 
higher cross sections. (ii) By measuring
appropriate observables one can get additional information on
the SUSY parameters. (iii) One can reduce the background.
We perform our calculations in the Minimal Supersymmetric
Standard Model (MSSM) \cite{nilles} and study the SUSY
parameter dependence of cross sections and decay rates.

\section{Production of Charginos and Neutralinos}

The masses and couplings of neutralinos and charginos are
determined by their mixing matrices. The explicit
forms of the mixing matrices and their dependences on the SUSY
parameters are given, e. g., in \cite{chnold}. 
Chargino 
pair production proceeds via $\gamma$ and $Z$ exchange in the
$s$--channel and ${\tilde \nu}_e$ exchange in the $t$--channel. The
amplitude for $e^+ e^- \to {\tilde \chi}_i^0 {\tilde \chi}_j^0$
has contributions from $Z$
exchange in the $s$--channel and ${\tilde e}_L$ and ${\tilde e}_R$ 
exchanges in the
$t$-- and $u$--channel. As an example we show in Fig. 1 contour
plots of the 
cross section of $e^+e^- \rightarrow {\tilde\chi}^0_1 {\tilde\chi}^0_2$ as
a function of the longitudinal electron and positron beam
polarizations $\cP_{-}$ and $\cP_{+}$, at
$\sqrt{s}=230$~GeV. The SUSY parameters we have chosen as
$M_2 = 152$~GeV, $\mu = 316$~GeV, $\tan\beta = 3$. The
corresponding neutralino masses are $m_{\tilde\chi^0_1} = 71$~GeV
and $m_{\tilde\chi^0_2} = 130$~GeV. The masses of the exchanged
selectrons we have taken as $m_{\tilde e_R} = 132$~GeV and 
$m_{\tilde e_L} = 176$~GeV in Fig. \ref{fig_1}\,a, whereas in 
Fig. \ref{fig_1}\,b we have 
changed the value of $m_{\tilde e_L}$ into 
$m_{\tilde e_L} = 500$~GeV. As can be seen, the cross section
depends sensitively on the $e^-$ and $e^+$ beam polarizations.
In Fig. 1b the cross section for
left--polarized $e^-$ beams is much smaller than in Fig. 1a,
because the contribution of $\tilde e_L$ exchange is much smaller.
From the polarization 
dependence one can obtain information on the mixing states of the
neutralinos and on the masses of the exchanged selectrons.
We have also studied the cross sections of chargino
production, $e^+e^- \rightarrow \tilde\chi^+_i \tilde\chi^-_j$. By a
suitable choice of the $e^-$ and $e^+$ beam polarizations the
relative importance of the $s$--channel or $t$--channel contributions
can be enhanced. In this way information on the mixing states of
the charginos and the mass of the exchanged ${\tilde \nu}_e$ 
can be obtained. For further results on neutralino and chargino 
production we refer to \cite{chne,kal,kneur,chnew} and the
references therein.
{\setlength\unitlength{1cm}
\begin{figure}[h]
\begin{center}
\psfig{figure=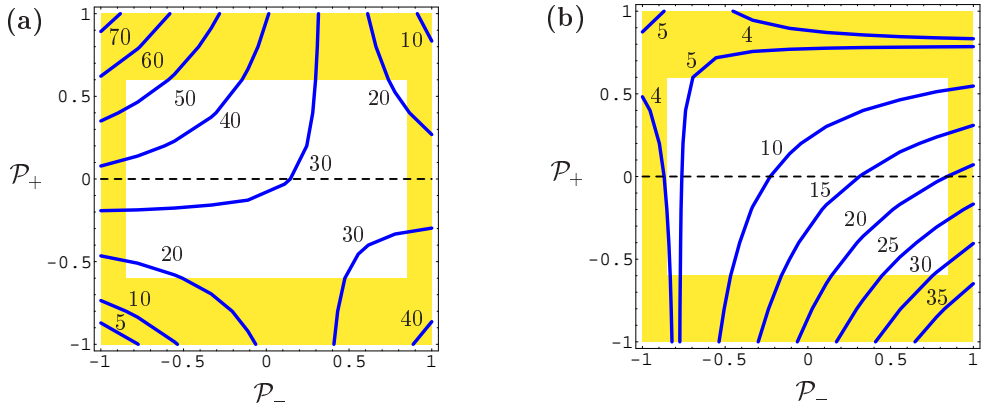,width=14cm}
\end{center}
\vspace{-5mm}
\caption{Contour lines of the cross section 
$\sigma(e^+e^-\to \tilde{\chi}^0_1 \tilde{\chi}^0_2)$ at 
$\sqrt{s} = 230$~GeV for $m_{\tilde{\chi}^0_1}=71$~GeV, 
$m_{\tilde{\chi}^0_2}=130$~GeV, the other parameters as in text.
The $e^-$ $(e^+)$ beam polarization is denoted by
$\cal P_-$ ($\cal P_+$). The white region is for $|\cal P_-|\le 
\mbox{85}\%$, $|\cal P_+|\le \mbox{60}\%$.
\label{fig_1}}
\end{figure}
}

\section{Sfermion Production}

In the discussion of third generation sfermions various aspects
of left--right mixing are important. For each sfermion of
definite flavour 
the interaction states $\ti f_L$ and $\ti f_R$ 
are mixed by Yukawa terms. 
Left--right mixing of the sfermions is described by the 
symmetric $2 \times 2$ mass matrices which depend on the soft
SUSY--breaking mass parameters 
$M_{\ti Q}$, $M_{\ti U}$ etc., and the trilinear scalar coupling 
parameters $A_t$, $A_b$ etc. (for details see
\cite{guha,sf,sfold}). 
The mass eigenstates are
$\ti f_1^{} = \ti f_L^{} \cth_{\ti f} + 
\ti f_R^{}\sin\theta_{\ti f}$,
$\ti f_2^{} = \ti f_R^{} \cth_{\ti f} - 
\ti f_L^{}\sin\theta_{\ti f}$, 
with $\theta_{\ti f}$ the sfermion mixing angle.
Strong $\ti f_L - \ti f_R$ mixing is expected for the third
generation sfermions, because in this case 
the Yukawa couplings can be large. The 
experimental search for the third generation sfermions will
be particularly interesting at an $e^+e^-$ linear collider,
where precise measurements of masses, cross sections and decay
branching ratios will be possible \cite{acco}. 
This will allow us to obtain information on 
the fundamental soft SUSY--breaking parameters. 

The reaction $e^+ e^- \ra \ti f_i \bar{\ti f_j}$ proceeds
via $\gamma$ and $Z$ exchange in the $s$--channel. The $\ti f_i$
couplings depend on the sfermion mixing angle $\theta_{\ti f}$.
In 
Figs. \ref{fig:stop11pol}\,a,\,b 
we show the contour lines of
the cross section $\sigma (e^+e^- \ra \ti t_1 \bar{\ti t_1})$ as 
a function of the $e^-$ and $e^+$ beam polarizations $\cP_-$ and 
$\cP_+$ at $\sqrt{s}=500$~GeV for 
$m_{\tilde t_1} = 200$~GeV and two values of $\cst$: $\cst = 0.4$ 
in (a) and $\cst= 0.66$ in (b). We have included initial--state
radiation (ISR) and SUSY--QCD corrections (for details see
\cite{sf,sfold}).
The white windows show the range of polarizations
$|\cP_{-}| < 0.9$ and $|\cP_{+}| < 0.6$.
As one can see, one can significantly increase the cross section by 
using the maximally possible $e^-$ {\it and} $e^+$ polarization. 
Moreover, beam polarization strengthens the $\cst$ dependence
and can thus be essential for determining the mixing angle. 
We have also calculated the cross sections for the production
of sbottoms, staus and $\tau$--sneutrinos. The results are in
\cite{sf}.\\ 
{\setlength \unitlength{1mm}
\begin{figure}[h]
\vspace{-5mm}
\begin{center}
\psfig{figure=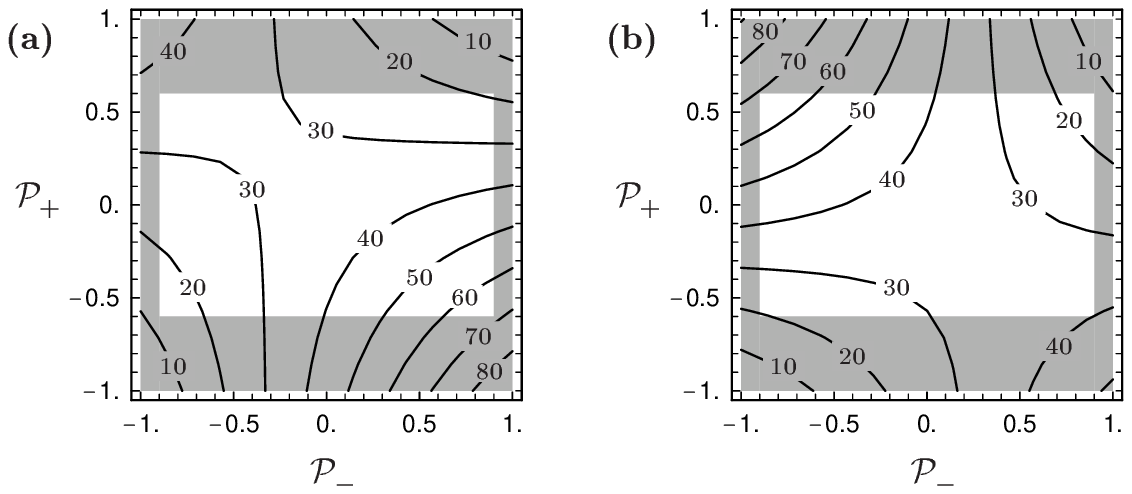,width=14cm}
\vspace{-5mm}
\end{center}
\caption{Dependence of $\s(e^+e^-\to\st_1\bar{\st_1})$ on 
degree of electron and positron polarization at $\sqrt{s}=500$ GeV, 
for $\mst{1}=200$ GeV, 
$\cst=0.4$ in (a) and $\cst=0.66$ in (b). 
\label{fig:stop11pol}}
\end{figure}
}
We have estimated the precision one may obtain for the
parameters of the $\st$ sector from cross section measurements.
We use the parameter point 
$\mst{1}=200$ GeV, 
$\cst=-0.66$ as an illustrative example: 
For 90\% left--polarized electrons (and unpolarized positrons) we have  
$\s_L(\st_1\bar\st_1)=44.88\;$fb, 
including SUSY--QCD, Yukawa coupling, and ISR corrections.  
For 90\% right--polarized electrons we have $\s_R(\st_1\bar\st_1)=26.95\;$fb.
According to the Monte Carlo study of \cite{nowak} 
one can expect to measure the $\st_1\bar\st_1$ production cross sections 
with a statistical error of $\D\s_L/\s_L = 2.1\,\%$ and 
$\D\s_R/\s_R = 2.8\,\%$ 
in case of an integrated luminosity of ${\cal L}=500\;\fbi$ 
(i.e. ${\cal L}=250\;\fbi$ for each polarization).
Scaling these values to ${\cal L}=100\;\fbi$ leads to 
$\D\s_L/\s_L = 4.7\,\%$ and $\D\s_R/\s_R = 6.3\,\%$.  
Figure~\ref{fig:pol-err}\,a 
shows the corresponding error bands and 
error ellipses in the $\mst{1}$--$\,\cst$ plane. 
The resulting errors on the stop mass and mixing angle are:  
$\D\mst{1}=2.2$ GeV, $\D\cst= 0.02$ for ${\cal L}=100\;\fbi$ 
and $\D\mst{1}=1.1$ GeV, $\D\cst= 0.01$ for ${\cal L}=500\;\fbi$. 
With the additional use of a 60\% polarized $e^+$ beam these values 
can still be improved by $\sim 25\%$.\\ 
{\setlength \unitlength{1mm}
\begin{figure}[h]
\begin{center}
\psfig{figure=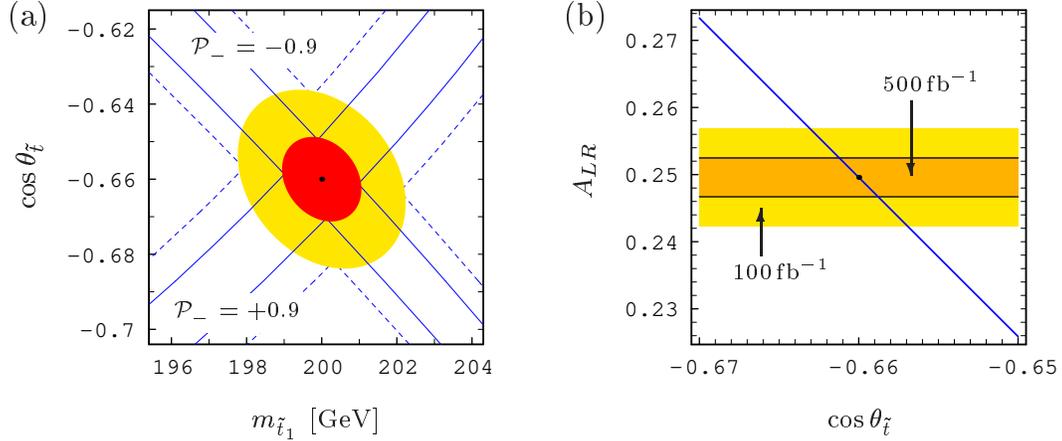,width=14cm}
\end{center}
\vspace{-2mm}
\caption{(a) Error bands and 68\% CL error ellipse for determining 
$\mst{1}$ 
and $\cst$ from cross section measurements;  
the dashed lines are for ${\cal L}=100\;\fbi$ and 
the full lines for ${\cal L}=500\;\fbi$. 
(b) Error bands for the determination of $\cst$ from $A_{LR}$.
In both plots $\mst{1}=200$ GeV, $\cst=-0.66$, $\sqrt{s}=500$ GeV, 
$\Pm=\pm 0.9$, $\Pp=0$.
\label{fig:pol-err}}
\end{figure}
}
For the determination of the mixing angle, one can also make use of 
the left--right asymmetry $A_{LR}$. 
At $\sqrt{s}=500$~GeV we get $A_{LR}(e^+e^-\to \st_1\bar{\st_1})=0.2496$ 
for the parameter point chosen 
and 90\% polarized electrons. 
Taking into account experimental errors as determined in \cite{nowak},  
a theoretical uncertainty of 1\%, and $\delta P/P=10^{-2}$ we 
get $\Delta A_{LR}=2.92\%$ (1.16\%) for ${\cal L}=100\,\fbi\;(500\,\fbi)$. 
This corresponds to $\Delta\cst=0.0031$ (0.0012). 
This is most likely the most precise method to determine the stop mixing 
angle. 
The corresponding error bands are shown in 
Fig.~\ref{fig:pol-err}\,b.

\section*{Acknowledgements}

A. B. is grateful to Prof. Tran Thanh Van and the organizers of
the IVth Rencontres du Vietnam for creating an inspiring
atmosphere at this conference. 
This work was supported in part by the ``Fonds zur F\"orderung der 
Wissenschaftlichen Forschung of Austria'', project no. P13139-PHY. 
W.P.~is supported by the Spanish ``Ministerio de Educacion y Cultura'' 
under the contract SB97-BU0475382, by DGICYT grant PB98-0693, and by the 
TMR contract ERBFMRX-CT96-0090.


\end{document}